\documentclass[doublecol]{epl2} 

\usepackage{stfloats}
\makeatletter
\def\hlinewd#1{%
  \noalign{\ifnum0=`}\fi\hrule \@height #1 \futurelet
   \reserved@a\@xhline}
\makeatother

\title{Coherence and incoherence collective behavior in financial market}
\shorttitle{Title} 

\author{Shangmei Zhao\inst{1} \and Qiuchao Xie\inst{1} \and Qing Lu \inst{5} \and Xin Jiang\thanks{Email:jiangxin@buaa.edu.cn}\inst{2,3} \and Wei Chen\inst{4}}
\shortauthor{Shangmei Zhao\etal}

\institute{
  \inst{1} School of Economics and Management, Beihang University, Beijing 100191, China \\
  \inst{2} LMIB $\&$ School of Mathematics and Systems Science, Beihang University, Beijing 100191, China \\
  \inst{3} Department of Engineering Sciences and Applied Mathematics, Northwestern University, Evanston, IL 60208, USA\\
  \inst{4} Center for Complex Network Research, Department of Physics, Northeastern University, Boston, MA 02115, USA\\
  \inst{5} China Export $\&$ Credit Insurance Corporation (Sinosure), Beijing 100033, China\\
}


\abstract{Financial markets have been extensively studied as highly complex evolving systems. In this paper, we quantify financial price fluctuations through a coupled dynamical system composed of phase oscillators. We find a Financial Coherence and Incoherence (FCI) coexistence collective behavior emerges as the system evolves into the stable state, in which the stocks split into two groups: one is represented by coherent, phase-locked oscillators, the other is composed of incoherent, drifting oscillators. It is demonstrated that the size of the coherent stock groups fluctuates during the economic periods according to real-world financial instabilities or shocks. Further, we introduce the coherent characteristic matrix to characterize the involvement dynamics of stocks in the coherent groups. Clustering results on the matrix provides a novel manifestation of the correlations among stocks in the economic periods. Our analysis for components of the groups is consistent with the Global Industry Classification Standard (GICS) classification and can also figure out features for newly developed industries. These results can provide potentially implications on characterizing inner dynamical structure of financial markets and making optimal investment tragedies.}

\begin{document}

\maketitle

\section{Introduction}

Financial markets, as typical highly complex evolving systems, have been attracting considerable scientific attention in recent complexity investigations\cite{mantegna2000introduction,sinha2010econophysics,stanley2002self}. One of the most studied financial market is the stock market, in which the dynamics of stock prices are believed to be characterized by collective and emergent behaviors\cite{gopikrishnan2001quantifying}. Interactions among the components of the stock market, which can be various stocks, manifest the internal structure of the system \cite{peron2012structure}. Statistical properties of the interactions have been widely studied by physicist and economists in order to understand the economy as a physical dynamical system and gain practical experience in asset allocation, portfolio-risk estimation and so on\cite{bouchaud2003theory}.

Much of the research in quantifying and interpreting the collective behavior in stock markets in the last decade has examined the correlations between price fluctuations of different stocks\cite{plerou2002random,tola2008cluster}. The correlation matrix, as a powerful tool in characterizing interactions, has been suggested to investigate many universal properties of different markets. For example, Kim and Jeong\cite{kim2005systematic} proposed improved methods to identify stock groups by analyzing the correlation matrix. Further, financial dynamics can also be analyzed through the cross-correlation matrix\cite{shen2009cross,shapira2014modelling}. Recently, many modern physical methods, such as random matrix theory (RMT)\cite{plerou2002random}, Brownian motion\cite{yura2014financial}, complex network approaches\cite{caldarelli2013reconstructing}, have been successfully introduced to study patterns of activity in the process of dynamic evolution of financial systems. A recent study by Peron and Rodrigues \cite{peron2011collective} shows that the emergence of collective behavior in stock-market networks can be analyzed by network synchronization, which has captured tremendous attention in network science\cite{arenas2008synchronization}.

It is generally believed that stock prices behave correlated collective dynamics in specific economic environments\cite{brunnermeier2008deciphering}, such as in financial crisis or debt crisis. Here we intend to quantify the correlated financial price fluctuations from the viewpoint of coupled dynamical systems. A coupled dynamical system is one composed of subsystems (or agents) with coupling\cite{kocarev1996generalized}. In the system, the states of certain agents can affect the time-evolution of others, which is quite analogous to the observed financial dynamics.

In this paper, we introduce a dynamical model composed of non-locally coupled phase oscillators. The coupling strength among oscillators is identified by the correlations among stocks. We show that the correlations induced coupling causes the oscillators split into two groups: one is composed of coherent, phase-locked oscillators, the other is composed of incoherent, drifting oscillators. This coherence and incoherence coexistence phenomena is also called the chimera state\cite{abrams2004chimera,abrams2008solvable}, which has greatly fascinated many nonlinear dynamic researchers.
The Financial Coherence and Incoherence (FCI) coexistence collective behavior provides a novel manifestation of the correlations between different stocks in the economic periods. During these periods, the size of the coherent groups, which corresponds to the number of a specific class of stocks, fluctuates according to real-world financial instabilities or shocks. For instance, the coherent group performs a rapid expansion during the $2008$ global crisis but shrinks a lot in the $9/11$ Attacks. Further, we also examine the involvement of each oscillator in the coherent groups in a series of time windows. Numerical results reveal that the stocks naturally fall into three groups: low coherent, middle coherent and high coherent group. The majority of stocks belonging to cyclical type sectors are in the high coherent group while those belonging to the defensive type fall into the low coherent group, which is consistent with the economic explanation. For stocks from information technology (IT) and related sectors, they mainly compose the middle coherent group, which indicates they may perform some new industry features.

\section{Coherence and incoherence dynamics in stock market}
The dynamics of daily correlation plays a pivotal role in many important applications in finance. In order to quantify correlations, we first calculate the return of stock $i=1,\ldots,N $, over a time scale $\Delta t$
\begin{equation}
\label{eq.1}
G_{i}(t)\equiv \ln P_{i}(t+\Delta t)-\ln P_{i}(t).
\end{equation}
Here $P_{i}(t)$ denotes the price of stock $i$ at time t. In this paper, we mainly deal with daily return time series and thus $\Delta t=1$ day. Since different stocks have varying levels of volatility (standard deviation), we introduce a normalized return
\begin{equation}
\label{eq.2}
g_{i}(t)\equiv\frac{G_{i}(t)-\langle G_{i}\rangle}{\sigma_i},
\end{equation}
where $\sigma_i \equiv \sqrt{\langle G_{i}^{2}\rangle-\langle G_{i}\rangle^2}$ is the standard deviation of $G_{i}$. Based on these, we can obtain the equal-time cross-correlation matrix $R$ with elements
\begin{equation}
\label{eq.3}
R_{ij} \equiv \langle g_{i}(t)g_{j}(t)\rangle.
\end{equation}
Notice here that the elements $R_{ij}$ are restricted to the domain $[-1,1]$, where $R_{ij}=1$ corresponds to perfect correlations, $R_{ij}=-1$ corresponds to perfect anti-correlations, and $R_{ij}=0$ corresponds to uncorrelated pairs of stock\cite{laloux1999noise,plerou1999universal}. To characterize the coupling strength between two stock return series, we introduce the coupling function between two stocks $i$ and $j$ as
\begin{equation}
C(i,j)=\sqrt{2*(1+R_{ij})}.
\end{equation}

This coupling kernel provides nonlocal positive coupling between various stocks during the price evolution process. It is assumed that $C_{ij}$ is close to $2$ if two stocks $i$ and $j$ prohibit quite similar price dynamics. While two stocks perform anti-correlation behaviors, the coupling is close to $0$. This definition is similar to the one introduced in \cite{mantegna1999hierarchical},which is used as a distance function. However, in this work, it is reasonable to assume the coupling is positively related with the correlation. Thus, we can obtain a symmetric $N \times N$ coupling matrix $C$, which presents a topology vision of the coupling in the stock price dynamics.

To obtain the coupling dynamics in a real-world stock market during a sufficient long time period, we use a database\footnote{http://www.optiontradingtips.com} consisting of the daily prices of Standard and Poor¡¯s $500$ stocks ($S\&P500$), a total number of $N=418$ stocks with full historical data from 3rd January $2000$ to 2nd November $2012$ are selected. We use $3230$ closures prices to calculate the daily return of each stock. Since the correlation between stocks varies in different time scales, it is natural to consider the time evolution of these respective coupling matrices $C(t)$. We set the time scale as quarterly financial period, that is $\Delta window = 62$ days. $\Delta window$ can be regarded as a time window and by moving $\Delta window$ forward by a setting time step $\delta t =1$ day, a series of coupling matrix is obtained respectively. Since we have $3229$ returns for each stock, a total amount of $M=3168$ matrices is obtained to study the collective dynamics of stocks evolving with time.

\begin{figure}
\centerline{
\includegraphics[width=8cm]{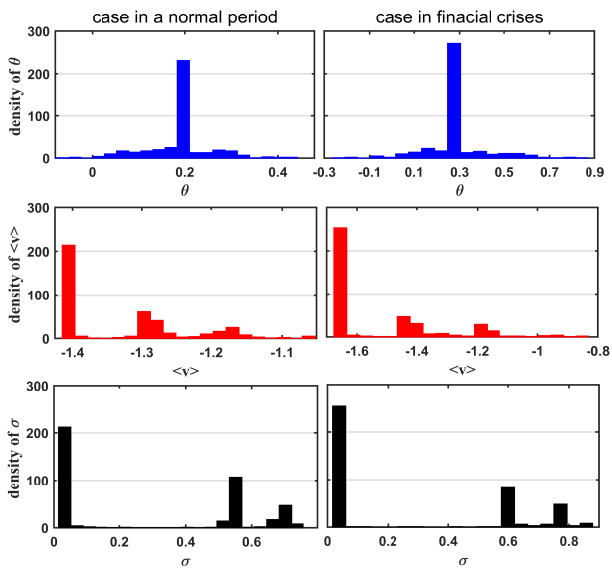}}
\caption{(Color online) Top panel (blue): distributions of the phase profile
 at snapshots of the stable states. Middle panel (red): corresponding distributions of the effective angular velocities $\langle v \rangle$. Bottom panel (black): corresponding distribution of the fluctuation of the instantaneous angular velocity of oscillators $\sigma$. All results are obtained after $10^4$ of iterations to ensure the system get into a stable state.}
\label{fig.1}
\end{figure}

To make a deep insight of the dynamics of oscillators induced by the coupling of financial market, we introduce here the Kuramoto and Battogtokh suggested model\cite{kuramoto2002coexistence} in a discrete form which reads
\begin{equation}
\label{mastereq}
\frac{d\theta_{i}}{dt}=\omega-\frac{1}{N}\sum_{j=1}^{N}C_{ij}\sin(\theta_{i}(t)-\theta_{j}(t)+\alpha), i=1,\dots,N.
\end{equation}
Here the scalar phase $\theta_{i}$ represents the state of the $i$th oscillator. The coupling strength is given by $C_{ij}$. Here we focus on the effect of $C_{ij}$ on the dynamics. For the other parameter $\omega$, since the stocks are chosen from the same stock market, we assume the intrinsic natural frequency $\omega$ are identical. In this sense, during our discussion, the frequency $\omega$ plays no role in the dynamics. In fact, if $\omega\neq 0$, one can transform $\omega=0$ without loss of generality by redefining $\theta=\theta+\omega t$ in Eq.~\ref{mastereq} without changing the form of Eq.~\ref{mastereq}. Throughout this work, we set $\alpha=\pi/2-0.10$ as a constant, which can be regarded as a measure of asymmetry of the distribution of phase differences between two stock price time series. We notice that a similar model is introduced in \cite{zhu2014chimera} to study chimera state on complex networks, where the coupling is topology depended.

We show there exist coherent (synchronized) and incoherent (irregular) patterns for oscillator dynamics in this system. Such coherence-incoherence pattern is observed simultaneously for each individual coupling matrix $C$. For $C$ in normal period or financial crises, the coherence-incoherence pattern differs a lot on the size. To illustrate this, we first give the density distribution of a snapshot of phases of oscillators after the system has involved into a steady state for two selected coupling matrix $C$ (see top panel, Fig.~\ref{fig.1}). We also consider the distribution of corresponding effective angular velocity of each oscillator $i$, which is defined as $\langle v_{i} \rangle = \lim_{T\rightarrow\infty}\frac{1}{T}\int_{t_{0}}^{t_{0}+T}\dot{\theta_{i}}dt$. The distribution of the fluctuation of the instantaneous angular velocity $\sigma_{i}$ of oscillator $i$ is also considered, where $\sigma_{i}^{2}= \lim_{T\rightarrow\infty}\frac{1}{T}\int_{t_{0}}^{t_{0}+T}(\dot{\theta_{i}}-\langle v_{i} \rangle)^{2}dt$.

\begin{figure}
\centerline{
\includegraphics[width=7.5cm]{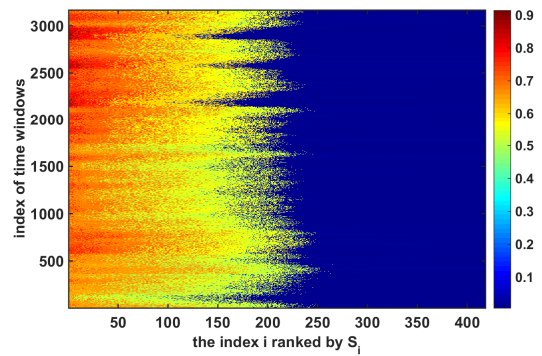}}
\caption{(Color online) Amplitude of the curve of $\sigma_{i}$, depicted as a contour plot in the index-time parameter space. Each $\sigma_{i}$ is calculated by averaging over $1000$ data points after the system is stable. Dark blue colors indicate zero amplitude.}
\label{fig.2}
\end{figure}

\begin{figure*}[bp]
\centerline{
\includegraphics[width=15cm]{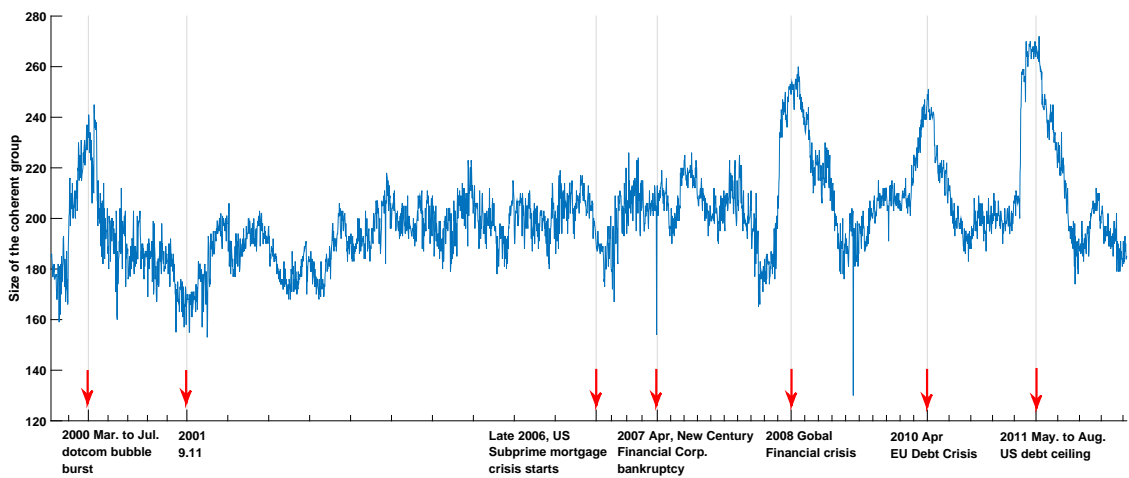}}
\caption{(Color online) The coherent group size varies in a long time economic cycle. The horizontal axis represents the time. The peaks and valleys correspond to several significant economic events during the period. }
\label{fig.3}
\end{figure*}

Numerical results are shown in Fig.~\ref{fig.1}. For the normal period case, $C$ is generated from the time window starting from $31st$ March to $29th$ June, $2004$. While for the financial crises case, $C$ is from $1st$ April to $29th$ June, $2010$, when the EU debt crises happened. In the top panel of Fig.~\ref{fig.1}, we show the distributions of instantaneous phases for both cases. One can find a distinct Dirac Delta distribution for locked oscillators and a Gaussian type distribution for the drifting ones, which indicates a quite conspicuous coherent and incoherent co-exist pattern. Comparing the left graph with right one, we show that in crises case, the size of coherent group is much larger than the one in normal case. The Dirac Delta pattern in the distribution of $\langle v \rangle$ presented in Fig.~\ref{fig.1} (middle) confirms that the oscillators in the coherent group have the same $\langle v \rangle$ value, which is around $-1.4$ and $-1.6$ respectively. On the other hand, the oscillators in the incoherent group perform a inhomogeneous profile of effective velocities. This is also evidenced by the peaks of the Dirac Delta distribution locating at $\sigma = 0$ in the distribution of the fluctuations of $\langle v \rangle$ (Fig.~\ref{fig.1}, bottom).

We observe that this coherent-incoherent pattern can be used to divide the oscillators into two groups. In the coherent group the oscillators are synchronized while the incoherent ones are drifting. In order to give a clear vision of the pattern, we need to rearrange the order of oscillators. To proceed, we first define the strength of global coupling for each index $i$ as $S_{i}=\sum_{j=1}^{N}C_{ij}$. This parameter indicates how strongly the selected stock $i$ coupled with others, which can be characterized as the financial influence of a stock. In this way, we label the oscillator with the smallest $S_{i}$ with index $1$ and ascend oscillators according to their individual value of $S_{i}$, which results a ranking of the previous indices. Since for each $C(t)$ there is a corresponding ranking of stocks, we obtain $M$ ranking sequences in total.

We demonstrate that the coherent group is mainly composed of stocks with relatively larger $S_{i}$ values and the coupling among these stocks can induce a synchronization dynamics of oscillators. While the incoherent group is composed of stocks with comparable smaller correlation strength and not strong enough to form a coherent collective behavior. The size of the coherent stock group fluctuates in this long time economic cycle. To identify the size of coherent stock group and also the corresponding stock indices in each time window efficiently, we consider the fluctuation of $\sigma_{i}$ for each oscillator. For oscillators in the coherent group, the corresponding $\sigma \approx 0$. Using this criteria, by setting a very small constant $\epsilon\ll1$, index $i$ is supposed to be in the coherent group once $\sigma_{i}<\epsilon$.
Fig.~\ref{fig.2} shows the results for the distribution of $\sigma$ in $M=3167$ time windows. For each time window, the sequence of $\sigma_{i}$ is pre-reordered by the individual global coupling strength $S_{i}$. As anticipated, we observe that the border between coherent and incoherent could be clearly identified by monitoring $\sigma_{i}$ in each $\Delta window$.

Using our criterion for the identification of coherent indexes group, we propose an algorithm to systematically calculate the indices set $Co(t)$ of coherent stock groups in $M$ numbers of time windows. For each $Co(t), t=1,\dots, M$, we start by checking from the index $i$ which corresponds to the largest $S_{i}$, that is, from left to right in the x-axis of Fig.~\ref{fig.2}. For each time window index $t$, for the reordered index sequence according to $S_{i}$,  while $\sigma_{i}<\epsilon$, $i$ is put into $Co(t)$. This checking procedure is terminated by the first $j$ encountered where $\sigma_{j}\geq\epsilon$. In this procedure, the incoherent indexes set $In(t)$ is obtained simultaneously. Actually, the power of the set $Co(t)$ is just the size of coherent group.

In Fig.~\ref{fig.3}, we visualize the time evolution of the coherent group size in a long time economic cycle. It is interesting to find that there exist several peaks for specific time periods. The first peak, from the left, locates at the period from March to July, $2000$, corresponds to the financial crisis when dotcom bubble burst in the US. The other three high peaks on the right side correspond to the global financial crisis start in $2008$, the EU Debt Crisis occurred in April $2010$ and the US debt ceiling between May and August $2011$, respectively. It is shown that during these financial crisis, much more oscillators tend to behave a coherent motion. However, for some unpredictable ¡°unusual¡± periods£¬ for example, the $9/11$ Attacks period, the size of the coherent group shrinks a lot. This pattern indicates that stock prices fails to react simultaneously for sudden artificial attacks and evolve in a much more independent way. We also point out that the coherence shrinks a lot as the subprime mortgage crises happened during late $2006$ and $2007$.
\begin{figure}
\centerline{
\includegraphics[width=6cm]{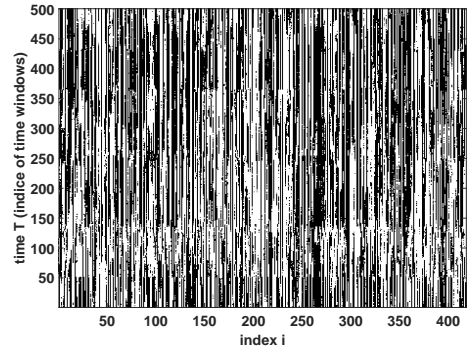}}
\caption{(Color online) Profile of the characteristic function $\chi(i,t)$ for all indexes varies with time $t$. Black colors indicate $0$ while white colors indicate $1$.}
\label{fig.4}
\end{figure}

\section{Inner structure detection of stock market by coherence dynamics}
As the time window proceeds forward, the members of the coherent group varies. For one stock, the state that whether it belongs to the coherent group or not is acting as a stochastic event with time. Considering the dynamics of a stock get involved in the coherent group or not during a long time, in Fig.~\ref{fig.4}, we visualize the evolution of $Co(t)$ and $In(t)$ by introducing a characteristic function $\chi(i,t)=1$ if $i\in Co(t)$ or $0$ if $i\in In(t)$. For each stock $i$, $\chi(i,t)$ is a time series with random Boolean values. The alternation of value $1$ and $0$ in the corresponding series indicates the dynamics of $i$ get involved into the coherent group and out of it. However, just from this direct illustration, one can hardly get any intuitive characteristics and features from the picture of $\chi(i,t)$ shown in Fig.~\ref{fig.4}.

\begin{figure}
\centerline{
\includegraphics[width=7.5cm]{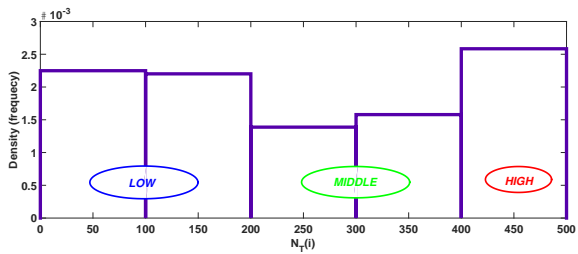}}
\caption{(Color online) Distribution of $N_{T}(i)$. The value of $N_{T}(i)$ ranges from $0$ to $500$. Three regions are mark with blue, green and red colors to identify the low, middle and high coherent respectively.}
\label{fig.5}
\end{figure}

\begin{figure}
\centerline{
\includegraphics[width=7cm]{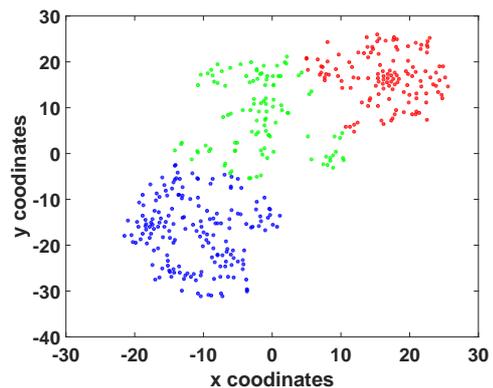}}
\caption{(Color online) A 2-dimension plot of $i$ clustering. Each $i$ is assigned with a pair of x-coordinate and y-coordinate calculated by the t-SNE method\cite{jain2010data}. The low coherent group is colored by blue dots, the middle coherent group is colored by green dots and the high coherent one is colored by red.}
\label{fig.6}
\end{figure}

\begin{table*}[bp]
\caption{Classification of grouped stocks in GICS sectors.  For each GICS sector (row), the percentages of stocks belong to high, middle and low coherent groups are given in column 2 to 4 respectively. In the last column, we identify whether the sector is cyclical or defensive.}
\begin{center}
\begin{tabular}[t]{|c|c|c|c|c|}
\hline
GICS sector & High coherent & Middle coherent & Low coherent & Cyclical or Defensive \\
\hlinewd{1pt}
Materials&	68.97\%&	13.79\%	&17.24\%	&Cyclical\\
Industrials&	60.00\%&	23.64\%&	16.36\%	&Cyclical\\
Consumer Discretionary &	49.25\% &	25.37\% &	25.37\% &	 Cyclical\\
Financial &	55.88\%&	17.65\%&	26.47\%&	Cyclical\\
\hline
Information Technology&	3.51\%&	82.46\%&	14.04\%&	 Cyclical\\
Telecommunications Services&	0.00\%&	66.67\%&	33.33\%&	 Defensive\\
\hline
Energy&	0.00\%	&2.78\%	&97.22\%&	Defensive\\
Utilities&	0.00\%&	3.45\%	&96.55\%	& Defensive\\
Consumer Staples&	7.14\%	&0.00\%	&92.86\%&	Defensive\\
Health Care&	2.38\%	&26.19\%	&71.43\%	& Defensive\\
\hline
\end{tabular}
\end{center}
\label{Table.1}
\end{table*}

In order to analyzed the essential collective pattern embedded in the $\chi(i,t)$ dynamics, we first consider the distribution of the total numbers $N_{T}(i)$ for $i$ get involved in the coherent group during the entire time period $T$, where $N_{T}(i)=\sum_{t=1}^{T}\chi(i,t)$. For the entire period, we calculate $N_{T}(i)$ for all the $418$ stocks. It is shown that the distribution of $N_{T}(i)$ has two crests and one trough which roughly divide the index set into three region: the LOW involved region, the MIDDLE involved one and the HIGH involved one (see Fig.~\ref{fig.5}). Based on this observation, we decided for the purpose of clustering the stock indexes by the k-means clustering method.

Notice here the matrix $\chi$ denotes a $M \times N$ Boolean value matrix, in which columns correspond to $N$ stock indexes and rows corresponds to $M$ dimensions. Each $\chi(i,t)$ is a high dimensional vector of dimension $M$. Clustering these high dimensional vectors directly is shown to be a quite difficult project which is also encountered when biologists trying to cluster gene sequences, since the operations on high dimensional vector is of high computational complexity. Besides, the visualization of them is also not easy. However, as the k-means method is essentially designed based on analyzing the distance between vectors, we can simplify this problem by using the Laplacian Eigenmaps (LE) dimensionality reduction technique which is introduced in \cite{belkin2003laplacian} to project $\chi(i,t)$ into the two-dimension plane while preserving the distance between the original vectors. In this sense, the visualization also becomes easy.

Actually we use the LE dimensionality reduction method to mapping the matrix $\chi$ in Fig.~\ref{fig.4} into $2D$ coordinates. Further, we execute the well-known k-means clustering method\cite{jain2010data} on this distance matrix. By setting $k=3$, one can determine three distinct clusters of the indexes. Fig.~\ref{fig.6} shows during the time period $[1,500]$, the $418$ stocks are divided into three groups identified by different colors. This result indicates that, for the entire time period in average, the stocks naturally can be divided into three groups: the low coherent, middle coherent and high coherent groups. Hence, for given time periods, we characterize each stock with the characteristic function $\chi(i,t)$ and determine which group it belongs to by a simple 3-means clustering procedure.

To demonstrate that the inner structure of the stock market by our method has some economical meanings, we study the classification of the groups of stocks from a traditional economic point of view. By inspecting the Global Industry Classification Standard (GICS) sector for each stock, we find the observed grouping of high, middle and low coherent has a quite direct economic explanation. In Tab. \ref{Table.1}, we list the percentages of stocks belong to these $3$ groups within each industry sector given by GICS. The majority of stocks from sectors of materials, industrials, consumer discretionary and financial are in the high coherent group. These business sectors are identified as cyclical by GICS. That is, they are quite sensitive to the business cycle, such that revenues of these industries are generally higher in periods of economic prosperity and expansion, and lower in periods of economic downturn and contraction. This may explain why these stocks evolve coherently during the business cycle. In contrast, stocks from energy, utilities, consumer staples and health care sectors fall into the low coherent group for they are identified as defensive. Sales and earnings of these defensive companies remain relatively stable during both economic upturns and downturns. Since they do not show much sensitivity to the business cycle, most of them evolve independently, or low coherently. For companies belonging to telecommunications services and information technology (IT), they mainly compose the middle coherent group, which indicates that corporations of IT type may still be a rapid developing economic components in the market. In fact, the industry development trend in this area diverges a lot during the business cycle.

\section{Conclusions}

In summary, we introduce a dynamical model composed of non-locally coupled phase oscillators to study the collective dynamics of stock markets.  We demonstrate that there exist coherence and incoherence coexistence collective behaviors in financial stock market.  The correlations among stocks make the system evolve into a stable state where the stocks split into two groups: one is composed of coherent, phase-locked oscillators, the other is composed of incoherent, drifting ones. We find that the fluctuations of the size of coherent group reflects the real-world financial instabilities or shocks quite well. By checking the time series of each oscillator getting involved in the coherent group during a long period, we show that the S\&P $500$ stocks can be well clustered into three groups: low coherent, middle coherent and high coherent group. It is demonstrated that the clustering results is well consistent with the GICS industry sector classification explanation. We argue that this method can provide nontrivial dynamical insights on the inner structure and evolution of financial markets. Considering nonidentical natural frequencies in this model will be our future work.

\acknowledgments
We thank the referees for helpful comments. This work is partially supported by the NSFC No. 11201017, 11290141, 71373017 and 11305219, Cultivation Project of NSFC (No. 91130019).

\bibliographystyle{eplbib}
\bibliography{ref}
\end{document}